\documentclass[letter]{aa}

\usepackage{graphicx}
\usepackage{txfonts}

\begin{document}

   \title{Planetary Nebulae distances in GAIA DR2}

   \author{S. Kimeswenger\inst{1,2}
   \and D. Barr\'ia\inst{1}
          }

   \institute{Instituto de Astronom{\'i}a, Universidad Cat{\'o}lica del Norte, Av. Angamos 0610, Antofagasta, Chile
   \and
Institut f{\"u}r Astro- und Teilchenphysik, Leopold--Franzens Universit{\"a}t Innsbruck, Technikerstr. 25, 6020 Innsbruck, Austria
}

   \date{Received 15 June 2018 / Accepted 17 July 2018}

% \abstract{}{}{}{}{}
% 5 {} token are mandatory

  \abstract
   {Planetary Nebula distance scales often suffer for model dependent solutions. Model independent trigonometric parallaxes have been rare. Space based trigonometric parallaxes are now available for a larger sample using the second data release of GAIA. }
   {We aim to derive a high quality approach for selection criteria of trigonometric parallaxes for planetary nebulae and discuss possible caveats and restrictions in the use of this data release.}
   {A few hundred sources from previous distance scale surveys were manually cross identified with data from the second GAIA data release (DR2) as coordinate based matching does not work reliable. The data are compared with the results of previous distance scales and to the results of a recent similar study, which was using the first data release GAIA DR1.}
   {While the few available previous ground based and HST trigonometric parallaxes match perfectly to the new data sets, older statistical distance scales, reaching larger distances, do show small systematic differences. Restricting to those central stars, were photometric colors of GAIA show a negligible contamination by the surrounding nebula, the difference is negligible for radio flux based statistical distances, while those derived from H$\alpha$ surface brightness still show minor differences. The DR2 study significantly improves the previous recalibration of the statistical distance scales using DR1/TGAS. }
  {}


   \maketitle
%
%-------------------------------------------------------------------

\section{Introduction}
The distances to planetary nebulae (PNe) always were facing the difficulty of lacking nearby targets, which can be reached well by direct methods. Trigonometric parallaxes have been obtained in a homogeneous long time line campaign by US Naval Observatory
\citep[USNO;][]{Harris2007} and from the Hubble Space Telescope \citep[HST;][]{Benidict09}.
Other studies \citep{HipparcosPN,Smith15} showed that Hipparcos spacecraft parallaxes seems to be not reliable. It was assumed that the contamination by the emission of the surrounding nebulae caused these problems.
Another model independent method for distances to PNe are a cluster membership as studied extensively by \citet{ClusterPNa}, \citet{ClusterPNb} and discussed in \citet{Frew16}.
Beside those model independent methods, a wide variety of statistical, model dependent individual distance scales have been derived. The most used ones certainly are those based on surface brightness versus angular sizes. Sometimes they include optical depth corrections.
All those methods have to be calibrated against a data set of nebulae with known distances. The older, widely used method is based on the 6 cm radio continuum flux, either using the ionized mass concept of \citet{Daub82} in the calibrations of \citet{CSK92} and \citet{SSV08} or, by means of the radio continuum brightness temperature as used by \citet{Steene_Zijlstra94} and calibrated with a galactic bulge sample.
The newest model given by \citet{Frew16} is based on similar ideas, but make use of the optical H$\alpha$ surface brightness and a wide set of various calibrators. Moreover they use a completely homogeneous data set for the brightness data derived by them self earlier \citep{Frew13}.
In \citet{Smith15} and in \citet{Frew16} a detailed description of the underlying physics and assumptions for all those methods is given.
With the upcoming of the GAIA project \citep{gaia_dr1} a new era was expected to start for many classes of objects. The first step into that was shown by \citet{Stang17} based on the combined TYCHO + GAIA DR1 solution called TGAS \citep{GAIA_DR1_TGAS}. With the second data release GAIA \citep[hereafter GDR2;][]{GAIA_DR2} now for the first time a complete homogeneous data set based only on GAIA is available. We present here the comparison of this new data set with common previous calibrations of PNe distances. Moreover we are comparing it to the preliminary TGAS results in \citet{Stang17}. Finally, we discuss possible caveats using the current GDR2.

\section{Sample Selection}
The trigonometric parallaxes of the USNO data set \citep{Harris2007} including 15 targets together to the HST parallaxes for 4 objects \citep{Benidict09} were selected. Based on the calibration of \citet{SSV08}, the catalogue of radio flux based distances by \citet[][table 1; hereafter SH10]{Stang10}
%(hereafter SH10)
was used as input catalogue and combined with the optical H$\alpha$ brightness based distances of \citet[][hereafter FPB16]{Frew16}. For the latter we use the distances form their Tab. 12. using the general "complete sample" calibration.
The GAIA data was taken interactively from the online data base 
using various search radii of up to 15\arcsec. The latter was necessary as often the coordinates in the catalogues for older extended PNe are not centered on the central star. We performed this for the 15 trigonometric parallaxes as well as for all 728 targets in SH10. We then compared the targets interactively in the Aladin Desktop Applet 
\citep{aladin}. We used various sky survey plate scans from SuperCOSMOS \citep{SuperCOSMOS} and 2MASS \citep{2MASS} and the help of finding charts for identification like those in the ESO Strasbourg Catalogue of Galactic PNe \citep{Acker92} and in the HASH PN database
\citep{HASH_PN}. However, in some cases of faint central stars even original literature like \citet{Weinberger77} and \citet{Kitter88} was used.
\begin{figure}[t!]
\centerline{\includegraphics[width=88mm]{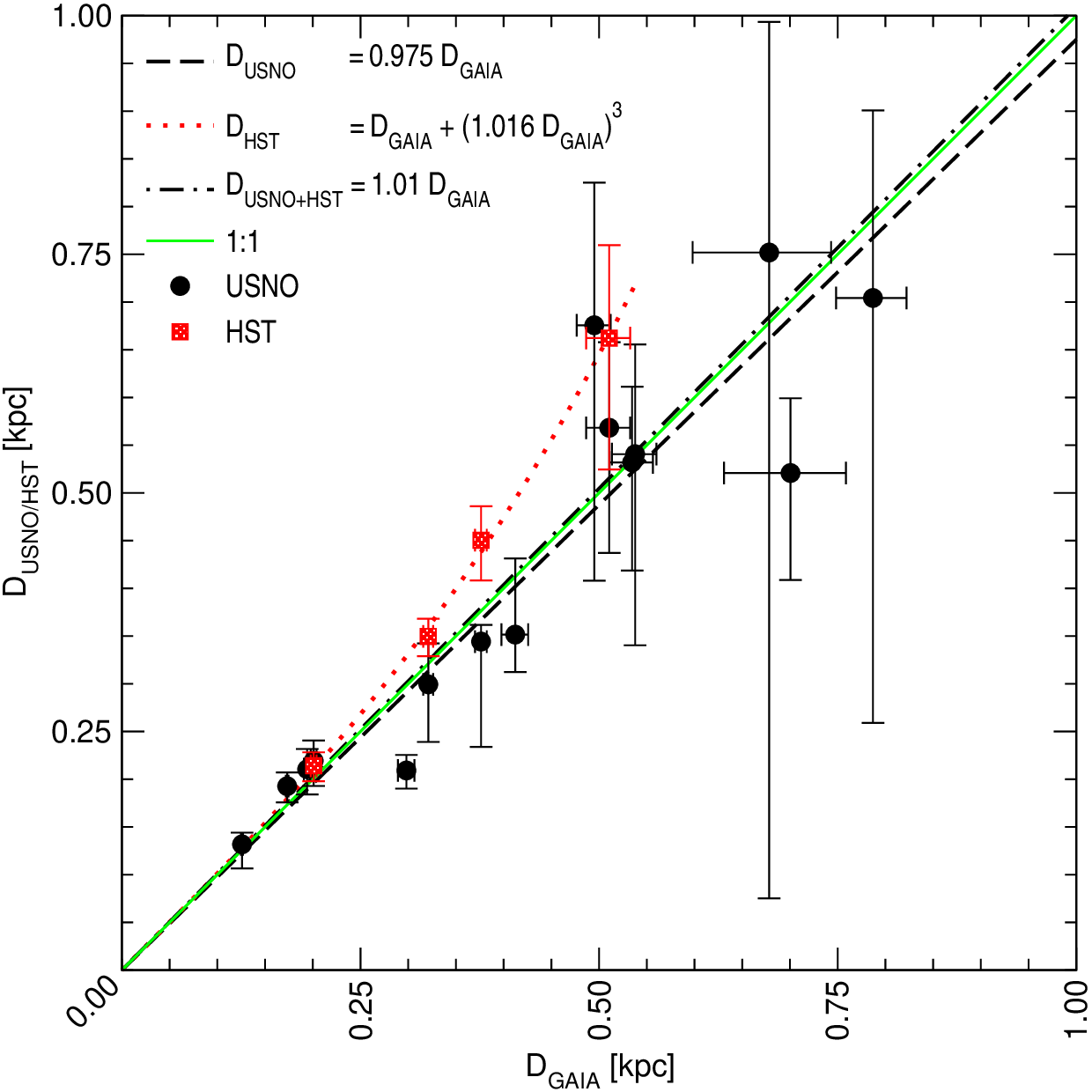}}
\caption{The USNO and HST trigonometric distances versus the GDR2 distances. The linear regressions show the USNO data set alone (dashed line) and the combined data set (dash-dotted line). The HST data set alone is best fitted by a relation adding a 3$^{\rm rd}$ order correction term (dotted curve). The thin green line gives the 1:1 relation.}
\label{fig_usno_hst}
\end{figure}
\relax
\noindent Only targets with a secure identification match, also in order of magnitude and clear identification on finding charts were used as positive matches. In total 382 out of the 728 sources from SH10 were matched. At two targets for the first time, the blue central source was found.
The complete data set with the cross identifications is given as electronic supplementary material only.
In the sample of 382 targets, many sources do have large formal (statistical) errors or even negative parallax values in the GAIA database. Only 199 sources do have a formal error of $\sigma_\pi / \pi < 0.50$ (out of which 170 are listed also in FPB16). As expected from the typical error for an isolated star of 0.03-0.04\,mas in the current data release \citep{GAIA_DR2_PARALAXES}, all sources beyond 4 kpc have errors above 0.25. Thus, we restricted our analysis to a sample below 4\,kpc which was splited into two sub-samples for $\sigma_\pi / \pi < 0.15$ (81 sources from SH10 with 75 also listed in FPB16) and for $\sigma_\pi / \pi < 0.25$ (32 sources from SH10 with 26 listed in FPB16).
The analysis in the remaining paper focuses on these two sub-samples, although we kept all sources in the online supplementary material to allow fast cross identification with future releases.

\section{Discussion}
\begin{figure*}[t!]
\sidecaption
\includegraphics[width=120mm]{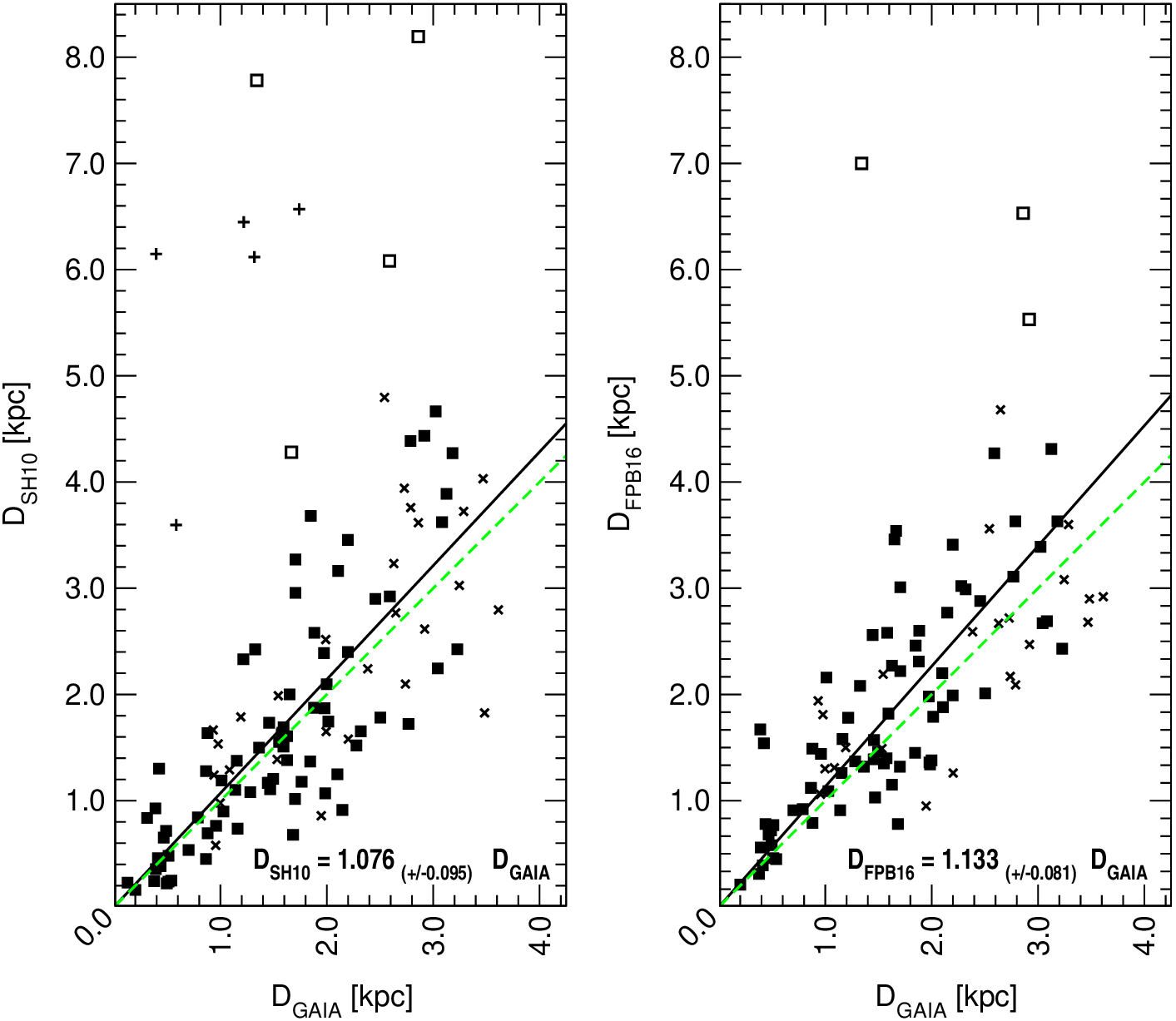}
\caption{
Comparison with statistical distance scales:
The catalogue of distances \citet{Stang10} (SH10) and of \citet{Frew16} (FPB16) are shown against GDR2 distances.
Squares ($\blacksquare$, $\square$) are the data for the targets with
$\sigma_\pi / \pi < 0.15$  in GAIA, where open
symbols indicate sources not used for the regression ($2.5\,\sigma$ filtering). The black solid lines gives the linear regression of these sources forced through the coordinate origin.
Crosses ($\times$) and plus ($+$) signs are used for targets with $0.15 \le \sigma_\pi / \pi < 0.25$, (latter not included for fitting), where sources mark with a plus are those filtered. The fitting including the sources with $\sigma_\pi / \pi > 0.15$ varies the result marginally within the line widths and thus is not shown separately for clarity. The green dashed lines give the 1:1 relation.}
\label{fig_stat_dist}
\end{figure*}
\relax

\subsection{Trigonometric Parallaxes}
\label{sec_trig}
The trigonometric parallaxes, obtained with nearly 20 years of ground based observations at USNO \citep{Harris2007}, are the up to now largest homogeneous reliable sample of that type \citep[see][]{Smith15,Frew16}. Four targets from that list have been also observed over a bit more than 3 years at HST \citep{Benidict09}. All targets from the USNO list, except PHL\,932, which was identified to be a compact H\,II region instead of a PN \citep{PHL932}, were used here and identified in GDR2. The parallaxes range from 1.3 to 8.0 mas ($\approx$0.1 to 0.8\,kpc) and have an average formal error of 3.2\% in the GDR2, but reaching up to 9\% for one source. The USNO data set is giving a mean error of 20\% and a maximum of 39\%, while the 4 additional sources observed at HST have a mean error of 10\% and a maximum of 17\%. At Fig.~\ref{fig_usno_hst} are shown the USNO and HST trigonometric distances vs. the GDR2 distances. Using only the USNO data, a quasi 1:1 match is found, while the HST data tends to a slightly longer distance scale for distances above 0.4\,kpc.
Although there was found a scaling factor of 1.17 (median) between the USNO and the HST parallaxes, \citet{Smith15} argued that due to the high scatter when comparing these two sets to each other, there is no difference in the distance scales within the statistical errors. However, the higher accuracy of the GAIA data does show now that the HST data in fact has some systematic deviation from the 1:1 relation.

\subsection{Statistical Distance Scales}
\begin{figure*}[ht]
\sidecaption
\includegraphics[width=120mm]{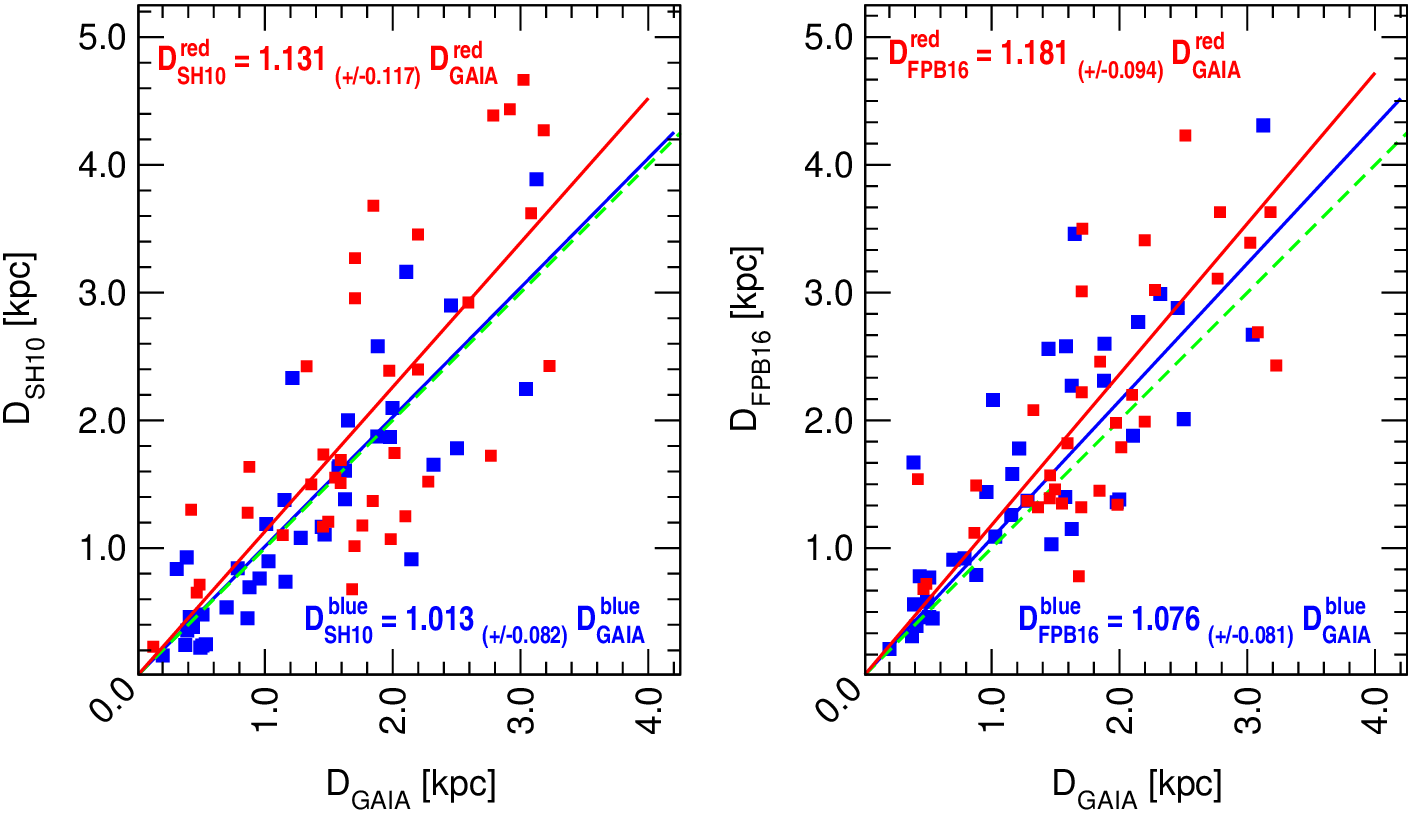}
\caption{Comparison with statistical distance scales using only sources with blue colors in GAIA $({\rm bp-rp}) < -0\fm25$ (fit line and symbols in blue) and the red objects (fit line and symbols in red). Symbols and lines are the same as in Fig.~\ref{fig_stat_dist}. The sources which were not used for fitting ($\sigma_\pi / \pi > 0.15$ and the clipped ones) are not shown for clarity.
}
\label{fig_stat_distBLUE}
\end{figure*}

As the trigonometric distances are restricted to nearby objects and to targets with large radii, and thus well-isolated central stars, we compared further on the data set to the matched sources in the two most commonly used statistical distance scales (SH10 and FPB16). For this purpose the sources were grouped into two sub-samples with parallax errors given in GAIA as being below 15\% (81 sources in SH10) and those with errors between 15\% and 25\%\footnote{This includes none of the peculiar objects from Sec 4.3.4 in FPB16.}. Regressions including weights for published statistical errors on individual sources were applied. Through a filtering process rejecting sources over $2.5\,\sigma$ (for the regression, see Fig.~\ref{fig_stat_dist}), we obtained a distance scale which is somewhat shorter by about 7-8\% compared to SH10 ($D_{\rm SH10}=1.076(\pm0.095)\,D_{\rm GAIA}; rms=0.53; R=0.80$) and about 13\% compared to FPB16 ($D_{\rm FPB16}=1.133(\pm0.081)\,D_{\rm GAIA}; rms=0.38; R=0.79$). While the first result do not show a statistically significant difference, the second one differs more than $1.6\,\sigma$ (Fig.~\ref{fig_stat_dist}).\\
On the other hand, the red filter of GAIA exhibit contamination by the H$\alpha$+[\ion{N}{ii}] emission lines \citep{GAIA_DR2_PHOTOMETRY}. Thus we find many objects with colors of (bp-rp) above zero (=red), even then if they show blue (B-V) colors (as the V band is not effected by the H$\alpha$+[\ion{N}{ii}] and the strong [\ion{O}{iii}] fall just between B and V into the low sensitivity region). We expect that central stars of PNe show $-0.65\le({\rm bp-rp})\le-0.25$. Hence we excluded in a further round all sources outside this expected color range to test quality of the parallaxes versus the photometric extraction of the pure stellar source (although, we know that we bias to larger older nebulae and exclude CSPNs with high extinction.)
We result then in very conservative samples of 47 and 45 targets for SH10 and FPB16, respectively.
The resulting comparison indeed remove any scaling factor with respect to the SH10 sample ($D_{\rm SH10}^{\rm blue}=1.013(\pm0.082)\,D_{\rm GAIA}^{\rm blue}; rms=0.47; R=0.87$) and has smaller differences with FPB16 ($D_{\rm FPB16}^{\rm blue}=1.076(\pm0.081)\,D_{\rm GAIA}^{\rm blue}; rms=0.31; R=0.81$), as we show in Fig.~\ref{fig_stat_distBLUE}. For comparison, the fit using only the red sequence targets give systematically higher inclinations where $D_{\rm SH10}^{\rm red}=1.131(\pm0.117)\,D_{\rm GAIA}^{\rm red}; rms=0.54; R=0.72$ and $D_{\rm FPB16}^{\rm red}=1.181(\pm0.094)\,D_{\rm GAIA}^{\rm red}; rms=0.39; R=0.75$. Further data over longer timescales might overcome part of these restrictions. Some targets have still a low number of  ({\tt visibility\_periods\_used$<$10}) compared to the 15 to 20 for most sources in the GDR2 catalogue and the limit of 8 to be used at all in the catalogue.

\subsection{Radio and optical surface brightness versus radius: \protect{\newline}GAIA DR2 vs. GAIA DR1/TGAS}
\begin{figure}[ht]
\centerline{\includegraphics[width=88mm]{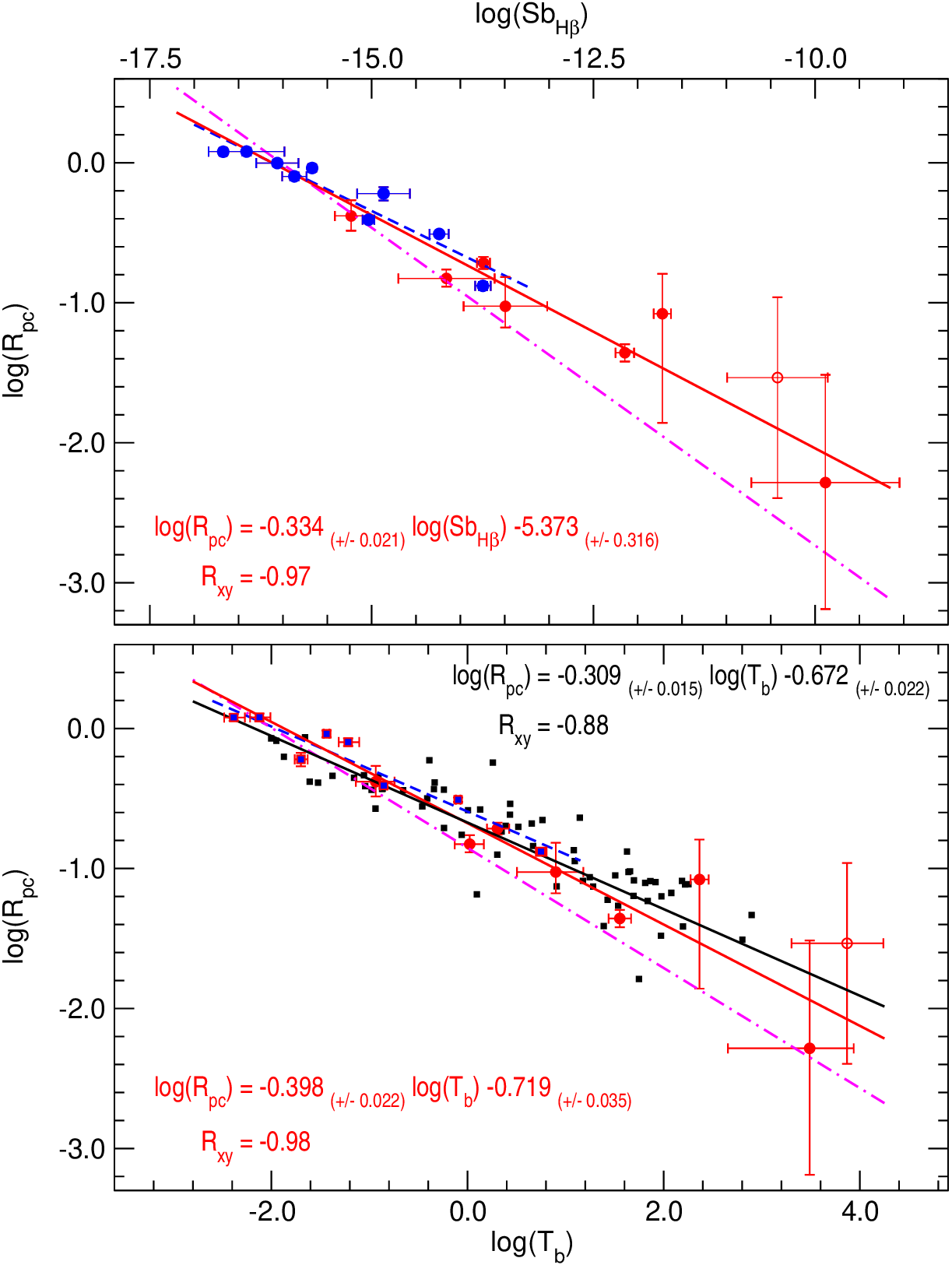}}
\caption{Comparison with calibration in \citet{Stang17}. Red symbols: targets only with TGAS parallaxes; blue symbols: the USNO targets; black symbols: our larger sample based on SH10.
Red fit lines are using identical targets as in the previous study, but using GDR2 distances. The black fit line uses the complete sample based on SH10.
SaSt~2-12 is marked as an open symbol.
The dash-dotted magenta lines give the fits by \citet{Stang17}.
}
\label{fig_stang17}
\end{figure}

By using the parallaxes from the first GAIA data release (GAIA DR1), \citet{Stang17} obtained a preliminary re-calibration of the statistical distance scales using the surface brightness.
They presented a correlation based on a few sources within the TGAS \citep{GAIA_DR1_TGAS} solution, and combined it with the same USNO data set we used here. With these data sets they revisited the calibration schemes of H$\beta$ surface brightness versus radius (derived from parallaxes), and that one of radio brightness temperature versus radius. The first method is very similar to the ansatz of FPB16, who used H$\alpha$ instead. For comparison purposes, we used exactly the same sources to provide a maximum of compatibility from our new data set. Thus also, as they did, the halo PN~SaSt~2-12 was excluded from the fit.
As is shown in the upper panel of Fig.~\ref{fig_stang17}, the regression we find with GDR2 has about the same correlation coefficient, but a much lower inclination as they found. For the radio surface temperature method \citep{Steene_Zijlstra94} we have also derived the relation (lower panel of Fig.~\ref{fig_stang17}) using all our data from the sample selected from SH10 presented in the previous section.
Intentionally no restriction to the blue sample was used to avoid the bias.
The results from this full sample is nearly identical to that one including only the targets used in the previous study. FPB16 was not used, as recalculating from H$\alpha$ to H$\beta$, might suffer systematic errors from extinction correction.
The differences between our GDR2 calibrations of statistical distance scales and that one derived from TGAS are striking. Moreover, as already the tests in the previous section have showed, exist possible caveats of GDR2 parallaxes towards high surface brightness objects with nebular contamination.

\section{Conclusions}
We have studied, using a carefully, not only coordinate based cross identification, a selected sample of 382 galactic planetary nebulae from the sample of 728 sources in the distance scale study of \citet{SSV08} listed in \citet{Stang10}. The selected sample include only targets with a clear identification match to the second GAIA release data set. Out of those matched sources, 57 have formally negative parallax values and 39 have, although giving good photometric values, no astrometric solution/parallax at all. Out of these 96 sources, 86 do have extremely red colors as not expected even for the ionizing source of a young planetary nebula.\\
The study of the remaining sample shows a perfect match of the distance scale to previous ground based USNO parallaxes of \citet{Harris2007}, reducing now enormously the uncertainties. However a pure HST based data set by \citet{Benidict09} clearly suffers from a systematic high order error term at distances above 0.35\,kpc. All of those targets are having blue GAIA colors (0\fm65$ < $(bp-rp)$<$0\fm25).
The  comparison  with  the  statistical  distance  scale  from \citet{SSV08} show  a  nearly  perfect  match,  up  to the here studied distances of 4 kpc, if the lists are restricted to central stars revealing blue target colors as well. Red targets with strong nebular contamination on the other hand show some systematic scale factor and much larger scatter than that one the formal error given in the GDR2 database would assume. {For a detailed discussion of the formal error versus real errors, including systematic errors due to a software bug for bright sources (G$<$13) and the slightly negative zero point, one may refer to \citet{GAIA_DR2_ASTROMETRY} and \citet{GAIA_DR2_PARALAXES}.} We thus conclude that the released data underestimates systematically the errors for those objects with emitting circumstellar material or high brightness nebulosity. This certainly will also have some impact on the interpretation of GDR2 data in star forming regions and OB associations.
As pointed out already in FPB16, parallaxes generally suffer from the L-K bias \citep{L-K}. The correction given there, assumes a constant error $\sigma_\pi$ and thus a steady increase of the fractional error $f = \sigma_\pi / \pi$ at larger distances (we used a constant $f < 0.15$) and a constant space density. As both do not apply here, a detailed study with Monte Carlo simulations similar to those for stars  \citep[][]{GAIA_DR2_PARALAXES} has to be done. Nevertheless, that requires a model for the spatial distribution as well as a model of the discovery probability variations in the Galaxy. \\
A comparison with the identical sample used for a similar study with GAIA DR1/TGAS data by \cite{Stang17} shows strong systematic effect, leading us to the conclusion, that these solutions from TGAS should not be used for stars surrounded by strong emission nebulae. However, even our newly derived correlation, although it was not restricted to the better parallaxes, has to be used with care due to bias effects.

\begin{acknowledgements}
This work has made use of data from the European Space Agency (ESA) mission
{\it Gaia} (\url{https://www.cosmos.esa.int/gaia}), processed by the {\it Gaia}
Data Processing and Analysis Consortium (DPAC,
\url{https://www.cosmos.esa.int/web/gaia/dpac/consortium}). This research has made use of the SIMBAD database and the ALADIN applet, operated at CDS, Strasbourg, France. D.B. was supported by FONDO ALMA-Conicyt Programa de Astronom\'ia/PCI 31150001.
\end{acknowledgements}
\bibliographystyle{aa}
%\bibliography{GAIA_PN_distance_references}

\begin{thebibliography}{29}
\expandafter\ifx\csname natexlab\endcsname\relax\def\natexlab#1{#1}\fi

\bibitem[{{Acker} {et~al.}(1998){Acker}, {Fresneau}, {Pottasch}, \&
  {Jasniewicz}}]{HipparcosPN}
{Acker}, A., {Fresneau}, A., {Pottasch}, S.~R., \& {Jasniewicz}, G. 1998, \aap,
  337, 253

\bibitem[{{Acker} {et~al.}(1992){Acker}, {Marcout}, {Ochsenbein}, {Stenholm},
  {Tylenda}, \& {Schohn}}]{Acker92}
{Acker}, A., {Marcout}, J., {Ochsenbein}, F., {et~al.} 1992, {The
  Strasbourg-ESO Catalogue of Galactic Planetary Nebulae. Parts I, II.} (ESO,
  Garching, Germany, 1047 p., ISBN 3-923524-41-2)

\bibitem[{{Benedict} {et~al.}(2009){Benedict}, {McArthur}, {Napiwotzki},
  {Harrison}, {Harris}, {Nelan}, {Bond}, {Patterson}, \&
  {Ciardullo}}]{Benidict09}
{Benedict}, G.~F., {McArthur}, B.~E., {Napiwotzki}, R., {et~al.} 2009, \aj,
  138, 1969

\bibitem[{{Boji{\v c}i{\'c}} {et~al.}(2017){Boji{\v c}i{\'c}}, {Parker}, \&
  {Frew}}]{HASH_PN}
{Boji{\v c}i{\'c}}, I.~S., {Parker}, Q.~A., \& {Frew}, D.~J. 2017, in IAU
  Symposium, Vol. 323, Planetary Nebulae: Multi-Wavelength Probes of Stellar
  and Galactic Evolution, ed. X.~{Liu}, L.~{Stanghellini}, \& A.~{Karakas},
  327--328

\bibitem[{{Bonnarel} {et~al.}(2000){Bonnarel}, {Fernique}, {Bienaym{\'e}},
  {Egret}, {Genova}, {Louys}, {Ochsenbein}, {Wenger}, \& {Bartlett}}]{aladin}
{Bonnarel}, F., {Fernique}, P., {Bienaym{\'e}}, O., {et~al.} 2000, \aaps, 143,
  33

\bibitem[{{Cahn} {et~al.}(1992){Cahn}, {Kaler}, \& {Stanghellini}}]{CSK92}
{Cahn}, J.~H., {Kaler}, J.~B., \& {Stanghellini}, L. 1992, \aaps, 94, 399

\bibitem[{{Daub}(1982)}]{Daub82}
{Daub}, C.~T. 1982, \apj, 260, 612

\bibitem[{{Evans} {et~al.}(2018){Evans}, {Riello}, {De Angeli}, {Carrasco},
  {Montegriffo}, {Fabricius}, {Jordi}, {Palaversa}, {Diener}, {Busso},
  {Cacciari}, \& {van Leeuwen}}]{GAIA_DR2_PHOTOMETRY}
{Evans}, D.~W., {Riello}, M., {De Angeli}, F., {et~al.} 2018, ArXiv e-prints
  [\eprint[arXiv]{1804.09368}]

\bibitem[{{Frew} {et~al.}(2013){Frew}, {Boji{\v c}i{\'c}}, \&
  {Parker}}]{Frew13}
{Frew}, D.~J., {Boji{\v c}i{\'c}}, I.~S., \& {Parker}, Q.~A. 2013, \mnras, 431,
  2

\bibitem[{{Frew} {et~al.}(2010){Frew}, {Madsen}, {O'Toole}, \&
  {Parker}}]{PHL932}
{Frew}, D.~J., {Madsen}, G.~J., {O'Toole}, S.~J., \& {Parker}, Q.~A. 2010,
  \pasa, 27, 203

\bibitem[{{Frew} {et~al.}(2016){Frew}, {Parker}, \& {Boji{\v
  c}i{\'c}}}]{Frew16}
{Frew}, D.~J., {Parker}, Q.~A., \& {Boji{\v c}i{\'c}}, I.~S. 2016, \mnras, 455,
  1459

\bibitem[{{Gaia Collaboration} {et~al.}(2018){Gaia Collaboration}, {Brown},
  {Vallenari}, {Prusti}, {de Bruijne}, {Babusiaux}, \&
  {Bailer-Jones}}]{GAIA_DR2}
{Gaia Collaboration}, {Brown}, A.~G.~A., {Vallenari}, A., {et~al.} 2018, ArXiv
  e-prints [\eprint[arXiv]{1804.09365}]

\bibitem[{{Gaia Collaboration} {et~al.}(2016){Gaia Collaboration}, {Brown},
  {Vallenari}, {Prusti}, {de Bruijne}, {Mignard}, {Drimmel}, {Babusiaux},
  {Bailer-Jones}, {Bastian}, \& et~al.}]{gaia_dr1}
{Gaia Collaboration}, {Brown}, A.~G.~A., {Vallenari}, A., {et~al.} 2016, \aap,
  595, A2

\bibitem[{{Hambly} {et~al.}(2001){Hambly}, {MacGillivray}, {Read}, {Tritton},
  {Thomson}, {Kelly}, {Morgan}, {Smith}, {Driver}, {Williamson}, {Parker},
  {Hawkins}, {Williams}, \& {Lawrence}}]{SuperCOSMOS}
{Hambly}, N.~C., {MacGillivray}, H.~T., {Read}, M.~A., {et~al.} 2001, \mnras,
  326, 1279

\bibitem[{{Harris} {et~al.}(2007){Harris}, {Dahn}, {Canzian}, {Guetter},
  {Leggett}, {Levine}, {Luginbuhl}, {Monet}, {Monet}, {Pier}, {Stone},
  {Tilleman}, {Vrba}, \& {Walker}}]{Harris2007}
{Harris}, H.~C., {Dahn}, C.~C., {Canzian}, B., {et~al.} 2007, \aj, 133, 631

\bibitem[{{Kwitter} {et~al.}(1988){Kwitter}, {Jacoby}, \& {Lydon}}]{Kitter88}
{Kwitter}, K.~B., {Jacoby}, G.~H., \& {Lydon}, T.~J. 1988, \aj, 96, 997

\bibitem[{{Lindegren} {et~al.}(2018){Lindegren}, {Hernandez}, {Bombrun},
  {Klioner}, {Bastian}, {Ramos-Lerate}, {de Torres}, {Steidelmuller},
  {Stephenson}, {Hobbs}, {Lammers}, {Biermann}, {Geyer}, {Hilger}, {Michalik},
  {Stampa}, {McMillan}, {Castaneda}, {Clotet}, {Comoretto}, {Davidson},
  {Fabricius}, {Gracia}, {Hambly}, {Hutton}, {Mora}, {Portell}, {van Leeuwen},
  {Abbas}, {Abreu}, {Altmann}, {Andrei}, {Anglada}, {Balaguer-Nunez},
  {Barache}, {Becciani}, {Bertone}, {Bianchi}, {Bouquillon}, {Bourda},
  {Brusemeister}, {Bucciarelli}, {Busonero}, {Buzzi}, {Cancelliere},
  {Carlucci}, {Charlot}, {Cheek}, {Crosta}, {Crowley}, {de Bruijne}, {de
  Felice}, {Drimmel}, {Esquej}, {Fienga}, {Fraile}, {Gai}, {Garralda},
  {Gonzalez-Vidal}, {Guerra}, {Hauser}, {Hofmann}, {Holl}, {Jordan},
  {Lattanzi}, {Lenhardt}, {Liao}, {Licata}, {Lister}, {Loffler}, {Marchant},
  {Martin-Fleitas}, {Messineo}, {Mignard}, {Morbidelli}, {Poggio}, {Riva},
  {Rowell}, {Salguero}, {Sarasso}, {Sciacca}, {Siddiqui}, {Smart}, {Spagna},
  {Steele}, {Taris}, {Torra}, {van Elteren}, {van Reeven}, \&
  {Vecchiato}}]{GAIA_DR2_ASTROMETRY}
{Lindegren}, L., {Hernandez}, J., {Bombrun}, A., {et~al.} 2018, ArXiv e-prints
  [\eprint[arXiv]{1804.09366}]

\bibitem[{{Luri} {et~al.}(2018){Luri}, {Brown}, {Sarro}, {Arenou},
  {Bailer-Jones}, {Castro-Ginard}, {de Bruijne}, {Prusti}, {Babusiaux}, \&
  {Delgado}}]{GAIA_DR2_PARALAXES}
{Luri}, X., {Brown}, A.~G.~A., {Sarro}, L.~M., {et~al.} 2018, ArXiv e-prints
  [\eprint[arXiv]{1804.09376}]

\bibitem[{{Lutz} \& {Kelker}(1973)}]{L-K}
{Lutz}, T.~E. \& {Kelker}, D.~H. 1973, \pasp, 85, 573

\bibitem[{{Majaess} {et~al.}(2014){Majaess}, {Carraro}, {Moni Bidin},
  {Bonatto}, {Turner}, {Moyano}, {Berdnikov}, \& {Giorgi}}]{ClusterPNb}
{Majaess}, D., {Carraro}, G., {Moni Bidin}, C., {et~al.} 2014, \aap, 567, A1

\bibitem[{{Majaess} {et~al.}(2007){Majaess}, {Turner}, \& {Lane}}]{ClusterPNa}
{Majaess}, D.~J., {Turner}, D.~G., \& {Lane}, D.~J. 2007, \pasp, 119, 1349

\bibitem[{{Michalik} {et~al.}(2015){Michalik}, {Lindegren}, \&
  {Hobbs}}]{GAIA_DR1_TGAS}
{Michalik}, D., {Lindegren}, L., \& {Hobbs}, D. 2015, \aap, 574, A115

\bibitem[{{Skrutskie} {et~al.}(2006){Skrutskie}, {Cutri}, {Stiening},
  {Weinberg}, {Schneider}, {Carpenter}, {Beichman}, {Capps}, {Chester},
  {Elias}, {Huchra}, {Liebert}, {Lonsdale}, {Monet}, {Price}, {Seitzer},
  {Jarrett}, {Kirkpatrick}, {Gizis}, {Howard}, {Evans}, {Fowler}, {Fullmer},
  {Hurt}, {Light}, {Kopan}, {Marsh}, {McCallon}, {Tam}, {Van Dyk}, \&
  {Wheelock}}]{2MASS}
{Skrutskie}, M.~F., {Cutri}, R.~M., {Stiening}, R., {et~al.} 2006, \aj, 131,
  1163

\bibitem[{{Smith}(2015)}]{Smith15}
{Smith}, H. 2015, \mnras, 449, 2980

\bibitem[{{Stanghellini} {et~al.}(2017){Stanghellini}, {Bucciarelli},
  {Lattanzi}, \& {Morbidelli}}]{Stang17}
{Stanghellini}, L., {Bucciarelli}, B., {Lattanzi}, M.~G., \& {Morbidelli}, R.
  2017, \na, 57, 6

\bibitem[{{Stanghellini} \& {Haywood}(2010)}]{Stang10}
{Stanghellini}, L. \& {Haywood}, M. 2010, \apj, 714, 1096

\bibitem[{{Stanghellini} {et~al.}(2008){Stanghellini}, {Shaw}, \&
  {Villaver}}]{SSV08}
{Stanghellini}, L., {Shaw}, R.~A., \& {Villaver}, E. 2008, \apj, 689, 194

\bibitem[{{van de Steene} \& {Zijlstra}(1994)}]{Steene_Zijlstra94}
{van de Steene}, G.~C. \& {Zijlstra}, A.~A. 1994, \aaps, 108, 485

\bibitem[{{Weinberger}(1977)}]{Weinberger77}
{Weinberger}, R. 1977, \aaps, 30, 343

\end{thebibliography}

\end{document}